\documentclass[pra,twocolumn,floatfix,superscriptaddress]{revtex4-1}
\usepackage{dcolumn,amsmath}
\usepackage{graphicx}
\usepackage{bm}
\usepackage{amssymb}
\usepackage{multirow}

\usepackage{amsfonts}
\usepackage{amsmath}
\usepackage{cancel}
\usepackage[normalem]{ulem}
\usepackage{xcolor}
\begin{document}
\title{ Rotating and vibrating symmetric top molecule  RaOCH$_3$ in the fundamental $\mathcal{P}$, $\mathcal{T}$-violation searches}

\author{Anna Zakharova} \email{zakharova.annet@gmail.com} 
%
\affiliation{St. Petersburg State University, St. Petersburg, 7/9 Universitetskaya nab., 199034, Russia} 
\affiliation{Petersburg Nuclear Physics Institute named by B.P. Konstantinov of National Research Centre
"Kurchatov Institute", Gatchina, 1, mkr. Orlova roshcha, 188300, Russia}
\date{Received: date / Revised version: date}

\begin{abstract}{
We study the influence of the rotations and vibrations of the symmetric top RaOCH$_3$ molecule on its effectiveness as a probe for the $\mathcal{P}$ and $\mathcal{T}$-violating effects, such as the electron electric dipole moment (eEDM) and the scalar-pseudoscalar electron-nucleon interaction (Ne-SPS). The corresponding enhancement parameters $E_{\rm eff}$ and $E_{\rm s}$ are computed for the ground and first excited rovibrational states with different values of the angular momentum component $K$. For the lowest $K$-doublet with $v_\perp=0$ and $K=1$ the values are $E_{\rm eff}=47.647\,\mathrm{GV/cm}$ and $E_{\rm s}=62.109\,\mathrm{kHz}$. The results show larger deviation from the equilibrium values than in triatomic molecules.

} 
\end{abstract}
\maketitle

\section{Introduction}
\label{intro}

The powerful strategy of searching the New physics is the study of the violation of fundamental discrete symmetries, namely the spatial reflection ($\mathcal{P}$), the time reversal ($\mathcal{T}$), and the charge conjugation ($\mathcal{C}$) \cite{khriplovich2012cp}. While such violations are present in the Standard model \cite{schwartz2014quantum,particle2020review} thanks to the complex phases in the Cabibbo-Kobayashi-Maskawa (CKM) \cite{Cabibbo1963,KobayashiMaskawa1973} and Pontecorvo–Maki–Nakagawa–Sakata (PMNS)  \cite{Pontecorvo1957,MNS1962} mixing matrices, some of the corresponding effects, such as the electron electric dipole moment (eEDM), are considerably suppressed \cite{Fukuyama2012,PospelovRitz2014,YamaguchiYamanaka2020,YamaguchiYamanaka2021}. This makes a suitable background for possible manifestations of the physics Beyond the Standard model.

An attractive feature of the particle EDM searches is that they can be performed in experiments with the polar molecules \cite{sandars1967measurability,sushkov1978parity}. The same experiments allow us to put limit on  the $\mathcal{P}$, $\mathcal{T}$-odd scalar-pseudoscalar nucleon-electron interaction  \cite{ginges2004violations,PospelovRitz2014,ChubukovLabzowsky2016}. Recently it was shown that such interaction can be induced by the nucleon EDM and $\mathcal{P}$, $\mathcal{T}$ violating hadronic interactions \cite{Flambaum6,flambaum2020effects}. The measurement of the oscillations in time of this interaction may be used for searches of the axion Dark matter \cite{flambaum2020limits,roussy2021experimental}. The sensitivities of the molecular spectra to the effects of the fundamental symmetries violation can not be measured directly and must be obtained from the ab-initio molecular computations \cite{KozlovLabzowsky1995, titov2006d, Safronova2017}.
Other $\mathcal{P}$, $\mathcal{T}$-odd effects can be studied this way, such as the electron–electron interaction mediated by the axionlike particle \cite{stadnik2018improved,dzuba2018new,maison2021electronic,maison2021axion}, and the magnetic quadrupole moment \cite{FDK14, maison2019theoretical}.

The current limits on the eEDM and Ne-SPS were obtained with the diatomic molecules ThO \cite{baron2014order,ACME:18,DeMille:2001,Petrov:14,Vutha:2010,Petrov:15,Petrov:17} and  HfF$^{+}$ \cite{Cornell:2017,Petrov:18}. The experiment is based on the existence of the closely spaced opposite parity doublets in the spectrum of these molecules. Let us elucidate shortly the nature of the states of interest.

For a given absolute value of the projection of the electronic angular momentum on the molecular axis $\Omega$ there exists two states $|+\Omega\rangle$ and $|-\Omega\rangle$. Naively one may expect that these states correspond to two degenerate energy levels, however the interaction with the molecular rotation results in their split known as $\Omega$-doubling. For the $\mathcal{P}$, $\mathcal{T}$-symmetric Hamiltonian the stationary states must have definite parity. Because both $\mathcal{P}$ and $\mathcal{T}$ change the sign of $\Omega$ the stationary states should be,
\begin{equation}
|\pm\rangle=\frac{1}{\sqrt{2}}\Big(|+\Omega\rangle\pm |-\Omega\rangle\Big),
\end{equation}
The external electric field $\mathcal{E}$ (usually assumed to be directed along the laboratory $z$ axis) breaks $\mathcal{P}$ symmetry and the effective Hamiltonian, restricted to the doublet, can be written as,
\begin{equation}
\hat{H}_{E}=\begin{pmatrix}\frac{\Delta E}{2}&d_z \mathcal{E}\\ d_z \mathcal{E} & -\frac{\Delta E}{2}\end{pmatrix},
\end{equation}
where $d_z=\langle \Omega|\hat{d}_z|\Omega\rangle$ is the electric dipole moment. The eigenstates then become the superpositions of the initial $|\pm\rangle$ states and their eigenvalues experience are shifted which constitutes the well-known Stark effect. If the strength of the electric field is sufficiently high $\mathcal{E}\geq \frac{\Delta E}{d_z}$ the molecule polarization reaches maximum. Then the molecular spectrum becomes sensitive to the presence of the $\mathcal{P}$, $\mathcal{T}$-odd interactions. It is manifested in the energy difference of the levels with opposite values of the total angular momentum projection on the laboratory axis $z$ which we will denote as $M$,
\begin{equation}
E_{+M}-E_{-M}\simeq P(2E_{\rm eff} d_e + 2 E_{\rm s} k_s),
\end{equation}
where $d_e$ is the value of eEDM and $k_s$ is a coupling constant for Ne-SPS. Coefficient P reflects the degree of polarization that may not reach $100\%$, e.g. for the most of the levels in the YbOH molecule the efficiency is less than $50\%$ \cite{petrov2021sensitivity}.  If one knows the enhancement parameters $E_{\rm eff}$ and $E_{\rm s}$ then one can extract the values $d_e$ and $k_s$ from this energy splitting.

The same principle can be applied to other closely spaced parity doublets. The triatomic molecules with linear equilibrium configurations allow the transverse molecular vibrations in two perpendicular planes characterized by two vibrational quantum numbers $v_x$ and $v_y$. The superposition of the two vibrations can be also considered as a rotation of the bent molecule around its axis. Thus, we can describe the bending modes of such molecules with the vibrational quantum number $v_\perp=v_x+v_y$ and the rovibrational angular momentum $l_v=-v_\perp,-v_\perp+2,\ldots v_\perp$. As in case of the $\Omega$ doublets, the states with opposite values of $l_v$ form the opposite parity doublet, and the Coriolis interactions cause their splitting known as $l$-doubling. The magnitude of the $l$-doubling is typically much less that the values of the $\Omega$-doubling, therefore such molecules require much smaller external fields for the full polarization \cite{Kozyryev:17}.

This makes the triatomic molecules with the heavy atoms, such as RaOH and YbOH, a promising platform for the $\mathcal{P}$, $\mathcal{T}$-odd interaction searches. Another advantage of the triatomic molecules is the possibility of the laser cooling of the same species that possess the parity doublets \cite{Isaev_2017}. This was experimentally demonstrated for monohydroxide molecules \cite{kozyryev2017sisyphus,steimle2019field,augenbraun2020laser}. Radium containing molecules also experience an enhancement of the $\mathcal{P}$, $\mathcal{T}$-odd effects associated with the large octupole deformation of the nuclei \cite{auerbach1996collective,spevak1997enhanced}.

More complex polyatomic molecules possess a richer rovibrational spectrum and allow new types of the opposite parity doublets. For example, the molecules of the symmetric top type such as RaOCH$_3$ and YbOCH$_3$ may possess a nonzero value of the total angular momentum projection on the molecular axis $K$ even in the electronic ground states and without transverse vibrations. These molecules also admit laser-cooling \cite{isaev2016polyatomic, kozyryev2016proposal,kozyryev2019determination,augenbraun2021observation}. The corresponding parity doublets, known as $K$-doublets, have even smaller splittings than the $l$-doublets and, thus, require even smaller external fields for the full polarization. The possibility to search for the Schiff moment on the $^{225}$RaOCH$_3^+$ ion was studied in \cite{yu2021probing}. The values of $E_{\rm eff}$ for a number of the MOCH$_3$ molecules (including RaOCH$_3$) were obtained for the fixed equilibrium configuration in \cite{zhang2021calculations}. 

The values of the enhancement parameters $E_{\rm eff}$ and $E_{\rm s}$ are usually computed for the fixed equilibrium configuration. However, even in the ground state there is a quantum uncertainty in displacements of the atoms from the equilibrium. This is aggravated in the rotational and excited vibrational states that are planned to be used in the measurements. The question of the influence of the quantum vibrations on the sensitivity of the molecule was studied for the triatomic molecules in \cite{prasannaa2019enhanced,gaul2020ab,ourRaOH,zakharova2021rovibrational}. It has not been addressed yet for the symmetric top type molecules.

The aim of the present work is to determine the sensitivities of the RaOCH$_3$, the molecule of the symmetric top type, to the presence of the eEDM and Ne-SPS interaction taking into account the effects of the molecular rotation and vibration.

\section{Born-Oppenheimer approximation}
\label{Sec:BornOppenheimer}

Because the vibrational frequenciefs of the OCH$_3$ are much higher than Ra -- OCH$_3$ bond stretching and bending frequencies, we will neglect the deformations of the ligand. We used the geometry of the ligand similar to the one obtained in \cite{yu2021probing}. The dimensions are given in the Table \ref{tbl:LigandGeom}.

\begin{table}[h]
\small
  \caption{The ligand geometry}
  \label{tbl:LigandGeom}
  \renewcommand{\arraystretch}{1.5}
  \begin{tabular*}{0.48\textwidth}{@{\extracolsep{\fill}}ll}
    \hline\hline
$r(O-C)$ & $2.600\,\mathrm{a.u.}$\\
$r(C-H)$ & $2.053\,\mathrm{a.u.}$ \\
$\angle(O-C-H)$ & $110.73^\circ$ \\
\hline\hline
  \end{tabular*}
\end{table}

\begin{figure}[]
\centering
  \includegraphics[width=0.5\textwidth]{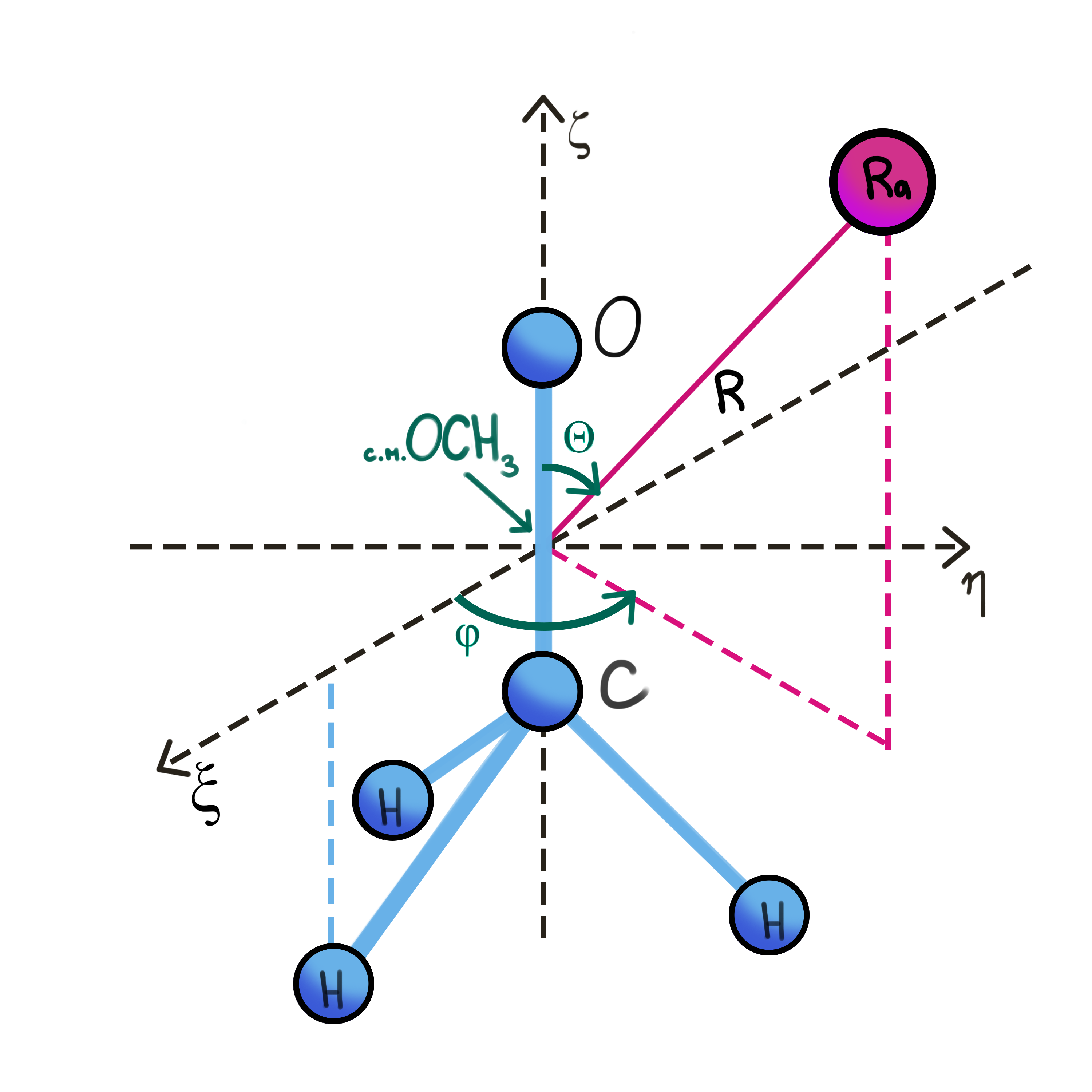}
  \caption{The RaOCH$_3$ molecule}
  \label{fig:molecule}
\end{figure}

We will employ the usual Born-Oppenheimer approximation, separating the total molecular wavefunction into a product of the electronic part and the part describing the motion of nuclei (which we will further call a nuclear wavefunction),
\begin{equation}
\Psi_{\rm total}\simeq \Psi_{\rm nuc}(R,\hat{R},\hat{r},\gamma)\psi_{\rm elec}(\{\vec{r}_i\}|R,\theta,\varphi),
\label{psiexp}
\end{equation}
where $R$, $\theta$ and $\varphi$ determine the geometry as shown on Fig.~\ref{fig:molecule}, $\hat{R}$ and $\hat{r}$ are the unit vectors in direction of the Ra -- ligand c.m. axis and ligand $\zeta$ axis (directed from C to O atom) correspondingly. The angle $\gamma$ determines the orientation of the CH$_3$ radical around $\zeta$ axis. $\psi_{\rm elec}$ is computed for the fixed molecular geometry $(R,\theta,\varphi)$.

The interaction of the electronic shell with the eEDM and the nuclei through the Ne-SPS can be described by $\mathcal{P}$, $\mathcal{T}$-odd effective Hamiltonian
\begin{align}
&\hat{H}_{\cancel{\mathcal{PT}}}=\hat{H}_d+\hat{H}_s^{(p)}+\hat{H}_s^{(n)},\\
&\hat{H_d}=  2d_e\sum_{i}
  \left(\begin{array}{cc}
  0 & 0 \\
  0 & \bf{\sigma_i E_i} \\
  \end{array}\right)\ 
 \label{Hd},\\
&\hat{H}_s^{(p)}=ik_s^{(p)}\frac{G_F}{\sqrt2}\sum_{j=1}^{N_{elec}}\sum_{I=1}^{N_{nuc}}{Z_I \rho_I\left(\vec{r_j}\right)}\gamma^0\gamma^5\\
&\hat{H}_s^{(n)}=ik_s^{(n)}\frac{G_F}{\sqrt2}\sum_{j=1}^{N_{elec}}\sum_{I=1}^{N_{nuc}}{N_I n_I\left(\vec{r_j}\right)}\gamma^0\gamma^5
\end{align}
where superscripts $(p)$ and $(n)$ denote the proton and neutron contributions correspondingly,
$G_F$ is Fermi constant, $Z_I$ is the proton number, $N_I$ is the neutron number, and $\rho_I$ is the charge density of the $I$-th nucleus normalized to unity, $n_I$ is the neutron density normalized to unity,  $\bf{E_i}$ is the inner molecular electric field acting on ith electron, $\bf{\sigma}$ are the Pauli matrices. As the open shell wavefunction (that determines the SCF value of the $\mathcal{P}$, $\mathcal{T}$-odd parameters) is concentrated near the Radium nucleus with the largest $Z_I$ and $N_I$ numbers, we will assume that the contribution from the other nuclei to $\hat{H}_s^{(p)}$ and $\hat{H}_s^{(n)}$ is small. We will also take the neutron density to be equal to the proton density, $n_{Ra}\simeq\rho_{Ra}$. In this approximation the proton and neutron contributions combine into,
\begin{equation}
\hat{H}_s^{(p)}+\hat{H}_s^{(n)}\simeq \hat{H}_s=ik_s\frac{G_F}{\sqrt2}\sum_{j=1}^{N_{elec}}{\rho_{Ra}\left(\vec{r_j}\right)Z_{Ra}\gamma^0\gamma^5},
\end{equation}
where we introduced $k_s$,
\begin{equation}
k_s=k_s^{(p)}+\frac{N_{Ra}}{Z_{Ra}}k_s^{(n)}.
\end{equation}
We use these definitions to be in accordance with the preceding computations in \cite{kudashov2014ab,gaul2019systematic,zakharova2021rovibrational,ourRaOH}, though one may expect the isoscalar $\sim Z_I+N_I$ and isotriplet $\sim Z_I-N_I$ components in Ne-SPS to be more natural. In principle the measurements with the different elements or, for high precision, even different isotopes of the same heavy element \cite{shitara2021cp} may allow to determine the nature of the interaction.

The sensitivity of the electronic shell in the given molecular configuration to these $\mathcal{P}$, $\mathcal{T}$-odd effects can be described by the parameters,
\begin{align}
E_{\rm eff}(R,\theta,\varphi)=\frac{\langle\psi_{elec}(R,\theta,\varphi)| \hat{H}_d|\psi_{elec}(R,\theta,\varphi)\rangle}{d_e{\rm sign}(\Omega)},\\ E_s(R,\theta,\varphi)=\frac{\langle\psi_{elec}(R,\theta,\varphi)| \hat{H}_s|\psi_{elec}(R,\theta,\varphi)\rangle}{k_s{\rm sign}(\Omega)}.
\end{align}
These parameters should be averaged over the rovibrational nuclear wavefunction (\ref{psiexp}):
\begin{align}
\label{Eeffaver}
E_{\rm eff}=\int dR d\hat{R} d\hat{r}d\gamma |\Psi_{nuc}(R, \hat{R}, \hat{r},\gamma)|^2 E_{\rm eff}(R,\theta,\varphi),\\
\label{Esaver}
E_s=\int dR d\hat{R} d\hat{r}d\gamma |\Psi_{nuc}(R,\hat{R}, \hat{r},\gamma)|^2 E_s(R,\theta,\varphi),
\end{align}

\section{Electronic computations}
\label{Sec:electronic}

To calculate the molecular orbitals by the Dirac-Harthree-Fock self-consistent field (SCF) method, as well as the potential surface with help of the coupled cluster method with single and double excitations (CCSD), we used a software package DIRAC 19. For the atoms composing the ligand i.e. O, C and H, we used the cc-pVTZ basis. To cut the costs of computations with heavy Radium atom we employed a 10-valence electron basis with a generalized relativistic effective core potential (GRECP) with spin-orbit interaction blocks \cite{titov1999generalized,mosyagin2010shape,mosyagin2016generalized}, developed by the Quantum Chemistry Laboratory of the PNPI \cite{QCPNPI:Basis}. This basis was used by us earlier in the computation of the $E_{\rm eff}$ and $E_{\rm s}$ parameters for the RaOH molecule \cite{ourRaOH}.

To compute the matrix elements of the $\mathcal{P}$, $\mathcal{T}$-odd parameters on the molecular orbitals we used the MOLGEP program, that corrects the behavior of the spinors obtained using GRECP in the core region with help of the method of one-center restoration based on equivalent bases \cite{Petrov:02,titov2006d,skripnikov2015theoretical}.

To obtain the values of the $E_{\rm eff}$ and $E_{\rm s}$ parameters on the CCSD level we applied the finite field method. In this approach the Hamiltonian is perturbed by the property $\hat{W}$ multiplied on a small parameter $\epsilon$,
\begin{equation}
    \hat{H}(\epsilon)\equiv\hat{H}+\epsilon\hat{W}
\end{equation}
Then the energy of the stationary state $|\psi\rangle$ is shifted by the expectation value of the property, multiplied on the $\epsilon$,
\begin{equation}
    E(\epsilon) = E + \epsilon\langle\psi| \hat{W}|\psi\rangle+O(\epsilon^2)
\end{equation}
This allows us to obtain the expectation values of the properties from the CCSD  energies, computed for the different perturbation parameters,
\begin{equation}
    \langle\psi|\hat{W}|\psi\rangle\simeq \frac{E(+\epsilon)-E(-\epsilon)}{2\epsilon}
\end{equation}

This technique could not be used straightforwardly within the DIRAC software because it allows only Kramers-restricted SCF computation with $\mathcal{T}$-even Hamiltonians and due to our use of the spinor-restoration procedure for the property matrix elements computations. However DIRAC does not rely on $\mathcal{T}$-symmetry in the CCSD computations. To circumvent its restrictions, we developed the program that modify the one-electron integrals with the matrix elements of the $\mathcal{P}$, $\mathcal{T}$-odd properties. The CCSD computations were then performed in DIRAC using the modified integrals. Previously this technique was successfully tested in our YbOH computations \cite{zakharova2021rovibrational}.

\section{Rovibrational wavefunctions}
\label{Sec:rovib}

The nuclear wavefunction can be obtained as an eigenstate of the nuclear Hamiltonian,
\begin{equation}
\hat{H}_{\rm nuc}\Psi_{\rm nuc}=E \Psi_{\rm nuc}
\end{equation}
In the present paper we restrict ourselves to the harmonic approximation in deviations from the equilibrium configuration. We will address the impact of the anharmonicities and non-adiabatic effects on the $\mathcal{P}$, $\mathcal{T}$-odd parameters for the symmetric top type molecules in the future work.

We will denote the body-fixed frame of reference axes as $X$, $Y$ and $Z$. The equilibrium configuration of the RaOCH$_3$ molecule corresponds to $\theta=0$ and $R=R_0$.

For the equilibrium configuration it is convenient to define the body-fixed frame of reference so that $X$, $Y$ and $Z$ coincide with the ligand principal axes $\xi$, $\chi$ and $\zeta$ correspondingly. Then they are also the principal axes of the whole molecule and the moment of inertia tensor is diagonalized,
\begin{equation}
I_{\rm tot}^{(eq)}=\begin{pmatrix}\mu R_0^2+I_\xi&0&0\\0&\mu R_0^2+I_\xi&0\\0&0&I_\zeta\end{pmatrix}
\end{equation}

For the non-equilibrium configuration we define the body-fixed frame of reference so, that the atom displacement would not contribute to the overall translations and rotations. For this the displacements $\vec{\delta}_k=\vec{r}_k-\vec{r}_{k,eq}$, where $\vec{r}_k$ is the coordinate of the $k$-th atom in the body-fixed frame of reference, should satisfy the Eckart conditions,
\begin{equation}
\sum_k m_k \vec{\delta}_k=0,\quad \sum_k m_k \vec{r}_{eq}\times \vec{\delta}_k=0
\end{equation}
As we keep the ligand to be rigid, the configuration of the molecule is determined by the coordinate of the Radium atom $\vec{r}_{Ra}$, the coordinate of the center of mass of the ligand $\vec{r}_{OCH3}$, and the Euler angles $\alpha, \beta, \gamma$ describing the orientation of the ligand. Namely, the ligand, which is at first oriented so that its axes $\xi$, $\chi$ and $\zeta$ coincide with the axes $X$, $Y$ and $Z$, is rotated by $\gamma$ around $Z$ axis, then by $\beta$ around $Y$ axis, and finally by $\alpha$ around $Z$ axis. The first Eckart condition then takes the form,
\begin{equation}
m_{Ra}\vec{\delta}_{Ra}+m_{OCH3}\vec{\delta}_{OCH3}=0,
\end{equation}

Defining $\vec{R}=\vec{r}_{Ra}-\vec{r}_{OCH3}$ and $\vec{\delta R}=\vec{R}-\vec{R}_{eq}$ we get,
\begin{equation}
\vec{\delta}_{Ra}=\frac{\mu}{m_{Ra}}\vec{\delta R},\quad \vec{\delta}_{OCH3}=-\frac{\mu}{m_{OCH3}}\vec{\delta R},
\end{equation}
where $\mu=\Big(\frac{1}{m_{Ra}}+\frac{1}{m_{OCH3}}\Big)^{-1}$ is the reduced mass of the Ra -- ligand system.

The second Eckart condition implies,
\begin{equation}
I\vec{\omega}+\mu\vec{R}_{eq}\times\frac{d}{dt}\vec{\delta R}=0,
\end{equation}
where $I$ is the ligand moment of inertia, and $\vec{\omega}$ is the angular velocity of the ligand in the body-fixed frame of reference.

We would like to apply this condition to the internal geometry variables $R$, $\theta$, $\varphi$ defined earlier and shown in Fig. \ref{fig:molecule}, and the orientation of the ligand $\alpha$, $\beta$, $\gamma$. Among these variables we can treat $\delta R=R-R_{0}$,  $\theta$ and $\beta$ as small parameters whereas the angles $\alpha$, $\gamma$ and $\varphi$ that specify the direction of the perturbation can be large. Then we obtain from the second Eckart condition,
\begin{equation}
\alpha=\varphi,\quad \gamma=-\varphi,\quad \beta=-\frac{\mu R_0^2}{I_{\xi}+\mu R_0^2}\theta.
\end{equation}
Because $\alpha+\gamma=0$, the displacement of the hydrogen atoms in the OCH$_3$ ligand remains to be small despite possible large values of the rotation angles.

Let us introduce three normalized variables,
\begin{equation}
q_{R}=\sqrt{\mu}\delta R,\quad q_x=\sqrt{\mathcal{I}}\theta \cos\varphi,\quad q_y=\sqrt{\mathcal{I}}\theta \sin\varphi,
\end{equation}
where
\begin{equation}
\mathcal{I}=\frac{\mu R_0^2 I_\xi}{\mu R_0^2+I_\xi}.
\end{equation}


Neglecting the centrifugial and Coriolis effects, the rovibrational Hamiltonian up to the second order in displacements takes the form,
\begin{align}
\hat{H}_{\rm nuc}\simeq &\frac{1}{2}\Big(\hat{\vec{J}}\cdot (I_{tot}^{(eq)})^{-1}\hat{\vec{J}}\Big)-\frac{1}{2}\sum_{k=R,x,y}\frac{\partial^2}{\partial q_k^2}\nonumber\\
&+V_{eq}+\frac{1}{2}\sum_{i,j=R,x,y}\frac{\partial^2 V}{\partial q_i\partial q_j}\Bigg\vert_{q_k=0}q_i q_j,
\end{align}

As we will see, the adiabatic potential $V(R,\theta,\varphi)$ only weakly changes with $\varphi$, and we can approximate it with the $\varphi$-averaged potential $\bar{V}(R,\theta)$. The symmetry of the molecule means that $\bar{V}(R,\theta)=\bar{V}(R,-\theta)$. All this means that at the harmonic approximation,
\begin{equation}
V(R,\theta,\varphi)\simeq V_{eq}+\frac{\omega_\parallel^2}{2} q_{R}^2+\frac{\omega_\perp^2}{2} (q_{x}^2+q_{y}^2),
\end{equation}
\begin{equation}
\omega_\parallel^2=\frac{\partial^2\bar{V}}{\partial q_R^2}\Bigg\vert_{q_k=0},\quad\omega_\perp^2=\frac{\partial^2\bar{V}}{\partial q_x^2}=\frac{\partial^2\bar{V}}{\partial q_y^2}\Bigg\vert_{q_k=0},
\end{equation}

Therefore, we obtained the Hamiltonian that is a sum of a rigid rotor with a moment of inertia $I_{tot}^{(eq)}$ and three decoupled harmonic oscillators. We can associate the vibrational quantum numbers $v_R$, $v_x$ and $v_y$ with $q_R$, $q_x$ and $q_y$ oscillators correspondingly. We will denote the total transverse vibrational quantum number as $v_\perp=v_x+v_y$.

The nuclear wavefunction then can be written as,
\begin{align}
\Psi_{\rm nuc}\simeq&\Psi_{\rm nuc}^{(0)}\equiv\psi_{JMK}(\alpha_{\rm m},\beta_{\rm m},\gamma_{\rm m})\nonumber\\
&\cdot\phi_{v_R}(\omega_\parallel,q_R)\phi_{v_x}(\omega_\perp,q_x)\phi_{v_y}(\omega_\perp,q_y)\label{Psinuc0}.
\end{align}

Here $\alpha_{\rm m}$, $\beta_{\rm m}$ and $\gamma_{\rm m}$ denote the Euler angles responsible for the body-fixed frame orientation with respect to the space-fixed frame. $\psi_{JMK}$ is the wavefunction of the rigid symmetric top rotor with definite square of the angular momentum $J(J+1)$, its projection $M$ on the space-fixed axis $z$, and projection $K$ on the body-fixed axis $Z$,
\begin{eqnarray}
\hat{J}^2\psi_{JMK}=J(J+1)\psi_{JMK},\\ \hat{J}_z\psi_{JMK}=M\psi_{JMK},\\
\hat{J}_Z\psi_{JMK}=K\psi_{JMK},
\end{eqnarray}

The functions $\phi_v(\omega,q)$ can be found to be the stationary wavefunctions of the Harmonic oscillator,
\begin{equation}
\phi_v(\omega,q)=\frac{1}{\sqrt{2^v v!}}\Big(\frac{\omega}{\Pi}\Big)^{\frac{1}{4}}\exp\Big(-\frac{\omega q^2}{2}\Big)H_v\Big(\sqrt{\omega}q\Big),
\end{equation}

Thus, the rough approximation for the averaged value of the property on a rovibrational state can be obtained with,
\begin{equation}
\langle E_{\rm eff,s}\rangle=\int d\alpha_{\rm m} d\beta_{\rm m} d\gamma_{\rm m} dq_R dq_x dq_y |\Psi_{\rm nuc}|^2 E_{\rm eff,s}(R,\theta,\varphi),
\end{equation}
where $E_{\rm eff,s}$ denotes the values of parameters obtained for the fixed molecular geometry.

\section{Impact of the $\varphi$-dependence of the potential}

For the approximated nuclear wavefunction \eqref{Psinuc0} only the $\varphi$-averaged value $\bar{E}_{\rm eff,s}$ contributes. To take into account the impact of the $\varphi$-dependence we use the first order perturbation theory. First we note that the equilibrium configuration of the RaOCH$_3$ molecule is symmetric under the transformations,
\begin{equation}
\varphi\mapsto -\varphi,\quad \varphi\mapsto \varphi+\frac{2\pi}{3},
\end{equation}
The same symmetry should be valid for the potential surface and $\mathcal{P}$, $\mathcal{T}$-odd parameters $E_{\rm eff,s}$. Therefore they can be decomposed into the Fourier series,
\begin{align}
V(R,\theta,\varphi)=&\bar{V}(R,\theta)+\delta V^{(1)}(R,\theta)\cos{3\varphi}\nonumber\\
&+\delta V^{(2)}(R,\theta)\cos{6\varphi}+\ldots,\\
E_{\rm eff,s}(R,\theta,\varphi)=&\bar{E}_{\rm eff,s}(R,\theta)+\delta E^{(1)}_{\rm eff,s}(R,\theta)\cos{3\varphi}\nonumber\\
&+\delta E^{(2)}_{\rm eff,s}(R,\theta)\cos{6\varphi}+\ldots,
\end{align}
For the purposes of our paper we truncated these series at $\cos{6\varphi}$ terms. Then to obtain the coefficients we require the values at $\varphi=0^\circ,30^\circ, 60^\circ$.

Let us treat the $\delta V(R,\theta,\varphi)=V(R,\theta,\varphi)-\bar{V}(R,\theta)$ as a small perturbation neglecting its dependence on $R$ (by setting $R=R_0$). The wavefunction then can be represented as,
\begin{align}
\Psi_{nuc}=&\Psi_{nuc}^{(0)}+\nonumber\\
&\psi_{JMK}(\alpha_{\rm m},\beta_{\rm m},\gamma_{\rm m})\phi_{v_R}(\omega_\parallel,q_R)\Phi_1(q,\varphi),
\end{align}
where $q=\sqrt{q_x^2+q_y^2}$ and $\Phi_1$ is the perturbation of the transverse vibration wavefunction. We decompose it into the Fourier series,
\begin{equation}
\Phi_1(q,\varphi)=\Phi_1^{(1)}(q)\cos{3\varphi}+\Phi_1^{(2)}(q)\cos{6\varphi}+\ldots
\end{equation}
with the constant term dropping out by orthogonality with $\Phi_0(q)=\phi_{0}(\omega_\perp,q_x)\phi_{0}(\omega_\perp,q_y)$ in $\Psi_{\rm nuc}^{(0)}$. The energy shift vanishes because $\Phi_0$ does not depend on $\varphi$,
\begin{align}
\delta E = &\int_0^{2\pi} d\varphi\int_0^{+\infty} dq\,q \Big[ |\Phi_0(q)|^2\delta V^{(1)}\cos{3\varphi}\nonumber\\
&+|\Phi_0(q)|^2 \delta V^{(2)}\cos{6\varphi}\Big]=0
\end{align}
The components relevant for our computation satisfy the equations,
\begin{align}
\Bigg[-\frac{1}{2q}\frac{\partial}{\partial q}\Big(q\frac{\partial}{\partial q}\Big)+\frac{9}{2q^2}+\frac{\omega_\perp^2}{2}q^2-\omega_\perp\Bigg]\Phi_1^{(1)}(q)\nonumber\\
=-\delta V^{(1)}\Big(R_0,\frac{q}{\sqrt{\mathcal{I}}}\Big)\Phi_0(q),\\
\Bigg[-\frac{1}{2q}\frac{\partial}{\partial q}\Big(q\frac{\partial}{\partial q}\Big)+\frac{36}{2q^2}+\frac{\omega_\perp^2}{2}q^2-\omega_\perp\Bigg]\Phi_1^{(2)}(q)\nonumber\\
=-\delta V^{(2)}\Big(R_0,\frac{q}{\sqrt{\mathcal{I}}}\Big)\Phi_0(q),
\end{align}
Interpolating $\delta V^{(n)}$ by a polynomial we solve the first equation in terms of the integrals of the rational functions of the Bessel functions, whereas the second one in terms of the integrals of the rational functions of the Whittaker functions. The integrals then are computed numerically.

The integration of $\cos{3n\varphi}$ products results in the following correction to the $\mathcal{P}$, $\mathcal{T}$-odd parameters due to the potential $\varphi$-dependence,
\begin{align}
\delta_{(\varphi)}E_{\rm eff,s}=&2\pi\int_{-\infty}^{+\infty}dq_R\int_0^{+\infty}dq\, q\,\phi_{0}(\omega_\parallel,q_R)^2\nonumber\\
&\cdot\Phi_0(q)\Bigg[\Phi_1^{(1)}(q)\delta E_{\rm eff,s}^{(1)}\Big(\frac{q_R}{\sqrt{\mu}},\frac{q}{\sqrt{\mathcal{I}}}\Big)\nonumber\\
&+\Phi_1^{(2)}(q)\delta E_{\rm eff,s}^{(2)}\Big(\frac{q_R}{\sqrt{\mu}},\frac{q}{\sqrt{\mathcal{I}}}\Big)\Bigg]
\end{align}

\section{The centrifugal and Coriolis effects}

When the centrifugal and Coriolis effects are taken into account the rovibrational kinetic energy for the RaOCH$_3$ molecule takes the form,
\begin{align}
T=&\frac{1}{2}(\vec{\Omega}\cdot I_{\rm tot}\vec{\Omega})+\Omega_z \zeta_{xy}^z\Big(q_x\dot{q}_y-q_y\dot{q}_x\Big)\nonumber\\
&+\Omega_x\zeta_{yR}^x\Big(q_y\dot{q}_R-q_R\dot{q}_y\Big)+\Omega_y\zeta_{Rx}^y\Big(q_R\dot{q}_x-q_x\dot{q}_R\Big)\nonumber\\
&+\frac{\dot{q}_R^2}{2}+\frac{\dot{q}_x^2+\dot{q}_y^2}{2}
\end{align}
where $\vec{\Omega}$ is the angular velocity of the body-fixed frame with respect to the space-fixed frame. The Coriolis coefficients are,
\begin{align}
\zeta_{xy}^z=-\zeta_{yx}^z=\frac{I_\xi}{\mu R_0^2+I_\xi},\\
\zeta_{yR}^x=-\zeta_{Ry}^x=\sqrt{\frac{I_\xi}{\mu R_0^2+I_\xi}},\\
\zeta_{Rx}^y=-\zeta_{xR}^y=\sqrt{\frac{I_\xi}{\mu R_0^2+I_\xi}}
\end{align}

The rovibrational Hamiltonian then takes the form \cite{watson1968simplification},
\begin{align}
\hat{H}=&\frac{1}{2}\Big(\hat{\vec{J}}-\hat{\vec{\pi}}\Big)\cdot \mathcal{M} \Big(\hat{\vec{J}}-\hat{\vec{\pi}}\Big)-\frac{1}{8}\mathrm{Tr}\,\mathcal{M}\nonumber\\
&-\frac{1}{2}\sum_{k=R,x,y}\frac{\partial^2}{\partial q_k^2}+V(R,\theta,\varphi)
\label{WatsonHamiltonian}
\end{align}
where $\hat{\vec{\pi}}$ is the vibrational angular momentum,
\begin{align}
\hat{\pi}_x=\zeta^x_{yR}\Big(-iq_y\frac{\partial}{\partial q_R}+iq_R\frac{\partial}{\partial q_y}\Big),\\
\hat{\pi}_y=\zeta^x_{Rx}\Big(-iq_R\frac{\partial}{\partial q_x}+iq_x\frac{\partial}{\partial q_R}\Big),\\
\hat{\pi}_z=\zeta^z_{xy}\Big(-iq_x\frac{\partial}{\partial q_y}+iq_y\frac{\partial}{\partial q_x}\Big),
\end{align}
and,
\begin{equation}
\mathcal{M}^{-1}=I_{\rm tot}+\mathcal{C},\quad \Big[\mathcal{C}\Big]_{\alpha\beta}=-\sum_{i,j,k=R,x,y}\zeta_{ik}^\alpha\zeta_{jk}^\beta q_i q_j
\end{equation}
The total moment of inertia may be decomposed into the series in the vibrational degrees of freedom,
\begin{equation}
I_{\rm tot}\simeq I_{\rm tot}^{(eq)}+I_{{\rm tot}}^{(1)} + I_{\rm tot}^{(2)}
\end{equation}
where $I_{\rm tot}^{(1)}$ is linear in $q_k$ and $I_{\rm tot}^{(2)}$ is quadratic in $q_k$. The tensor $\mathcal{M}$ then can be represented as,
\begin{align}
\mathcal{M}\simeq \Big(I_{\rm tot}^{(eq)}\Big)^{-1}-\Big(I_{\rm tot}^{(eq)}\Big)^{-1} I_{\rm tot}^{(1)} \Big(I_{\rm tot}^{(eq)}\Big)^{-1}\nonumber\\
+\Bigg(-\Big(I_{\rm tot}^{(eq)}\Big)^{-1} \Big(I_{\rm tot}^{(2)}+\mathcal{C}\Big) \Big(I_{\rm tot}^{(eq)}\Big)^{-1}\nonumber\\
+\Big(I_{\rm tot}^{(eq)}\Big)^{-1} I_{\rm tot}^{(1)} \Big(I_{\rm tot}^{(eq)}\Big)^{-1} I_{\rm tot}^{(1)} \Big(I_{\rm tot}^{(eq)}\Big)^{-1}\Bigg)
\end{align}

Because $\Big[I_{\rm tot}^{(eq)}\Big]_{XX}=\Big[I_{\rm tot}^{(eq)}\Big]_{YY}\gg \Big[I_{\rm tot}^{(eq)}\Big]_{ZZ}=I_\zeta$ we are primarily interested in the contribution to $\Big[\mathcal{M}\Big]_{ZZ}$. Neglecting the contributions from other components also allow us to preserve the factorization of $\Psi_{\rm nuc}$ into the product of the rotational and vibrational wavefunctions because then only $\hat{J}_Z$ component couples to the vibrational degrees of freedom. We can replace $\hat{J}_Z$ with its eigenvalue $K$. The neglected components of $\mathcal{M}$ give the centrifugal distortions due to the rotations of the molecule around axes $X$ and $Y$, and couplings between the transverse vibrations $q_x$, $q_y$, and the longitudinal mode $q_R$.

Then we obtain the following contribution from the first two terms in \eqref{WatsonHamiltonian} to the vibration Hamitlonian,
\begin{equation}
\delta_{\rm rot} \hat{H}=\omega_\perp\Delta(\omega_\perp)(q_x^2+q_y^2),
\end{equation}
where $K$-dependent correction to the $\omega_\perp$ depends on the eigenvalue of $l_v$,
\begin{align}
\omega_\perp\Delta(\omega_\perp)=&-\frac{4\tilde{K}^2-1}{8I_\zeta^2}\frac{(I_\xi-I_\zeta) \mu R_0^2+I_\xi^2}{I_\xi(\mu R_0^2+I_\xi)}\nonumber\\
&+\frac{(\zeta_{xy}^z)^2}{I_\zeta^2},\quad \tilde{K}=K-\zeta_{xy}^z l_v.
\label{omegaRot}
\end{align}
This represents the centrifugal effect - the larger is $K$, the smaller becomes the effective value of $\omega_\perp$ and, thus, the wider becomes the ground state.

The second term introduces the mixing between $q_x$ and $q_y$ modes due to the Coriolis force. However the operator $\hat{\pi}_z=\zeta_{xy}^z\hat{l}_v$, where $\hat{l}_v=-i\frac{\partial}{\partial\varphi}$. It commutes with the harmonic Hamiltonian for $q_x$ and $q_y$. Thus, these two operators have a common basis with eigenvalues of $l_v=-v_\perp,-v_\perp+2,\ldots v_\perp$. The ground state $v_x=v_y=0$ with the wavefunction $\Phi_0(q)$ happens to be also an eigenfunction of $\hat{\pi}_z$ with a zero eigenvalue because it does not depend on $\varphi$. Thus, for the vibrational ground state the effect of the Coriolis mixing vanishes.

\section{The excited vibrational states}

The similar analysis can be performed for the states with excited transverse modes $v_\perp>0$. Because of the Coriolis term the eigenstate should be an eigenfunction of $\hat{l}_v$. Because of the importance of the anharmonicities for the higher excited vibrational states we will restrict ourselves to $v=1$. The wavefunctions with $l_v=\pm 1$ then take the form,
\begin{align}
\Phi_{0,v_\perp=1,l_v=1}(q,\varphi)=\sqrt{\omega_\perp}{\pi}q e^{i\varphi}\exp\Big(-\frac{\omega_\perp q^2}{2}\Big),\nonumber\\
\Phi_{0,v_\perp=1,l_v=-1}(q,\varphi)=\sqrt{\omega_\perp}{\pi}q e^{i\varphi}\exp\Big(-\frac{\omega_\perp q^2}{2}\Big).
\end{align}
Just like with the ground state we can take into account the centrifugal effects using the correction \eqref{omegaRot} for the $\omega_\perp^2$.

The wavefunctions $\Phi_{0,v_\perp=1,l_v=\pm 1}$ contain only terms with $\cos\varphi$ and $\sin\varphi$. The product $\delta V(R,\theta,\varphi)\Phi_{0,v_\perp=1,l_v=\pm 1}$ no terms with $\cos{3\varphi}$ and $\cos{6\varphi}$ appear. Therefore, in the first-order perturbation theory no such term will appear in the correction to the wavefunction. Therefore no correction to the $\mathcal{P}$,$\mathcal{T}$-odd parameters appear due to the $\varphi$-dependence of the potential.

One may note that within the first-order perturbation theory in $\delta V(R,\theta,\varphi)$ the correction containing $\cos{3\varphi}$ or $\cos{6\varphi}$ may only appear when the $l_{v}=3N$, where $N$ is some integer number. In this case sort of a resonance happens between the $\varphi$-dependence of the wavefunction and of the $\mathcal{P}$,$\mathcal{T}$-odd parameter. The state sensitivity to the $\mathcal{P}$, $\mathcal{T}$-odd effects may be somewhat enhanced or decreased thanks to their $\varphi$-dependence. However for the lowest of such states $v_\perp=3$, $l_v=3$ we estimated that the correction to $E_{\rm eff}$ would be about $\sim \frac{\delta E_{\rm eff}^{(1)}(R_0,\theta_m)\delta V^{(1)}(R_0,\theta_m)}{\omega_\perp}\sim 10^{-3} \frac{\mathrm{GV}}{\mathrm{cm}}$, where $\theta_m$ is the maximum of the wavefunction $\Phi_{0,v_\perp=3,l_v=3}$. Hence, we will not study this effect in more detail in the present paper.

\section{Results and discussion}
\label{Sec:results}

\begin{figure}[h]
\centering
  \includegraphics[width=0.5\textwidth]{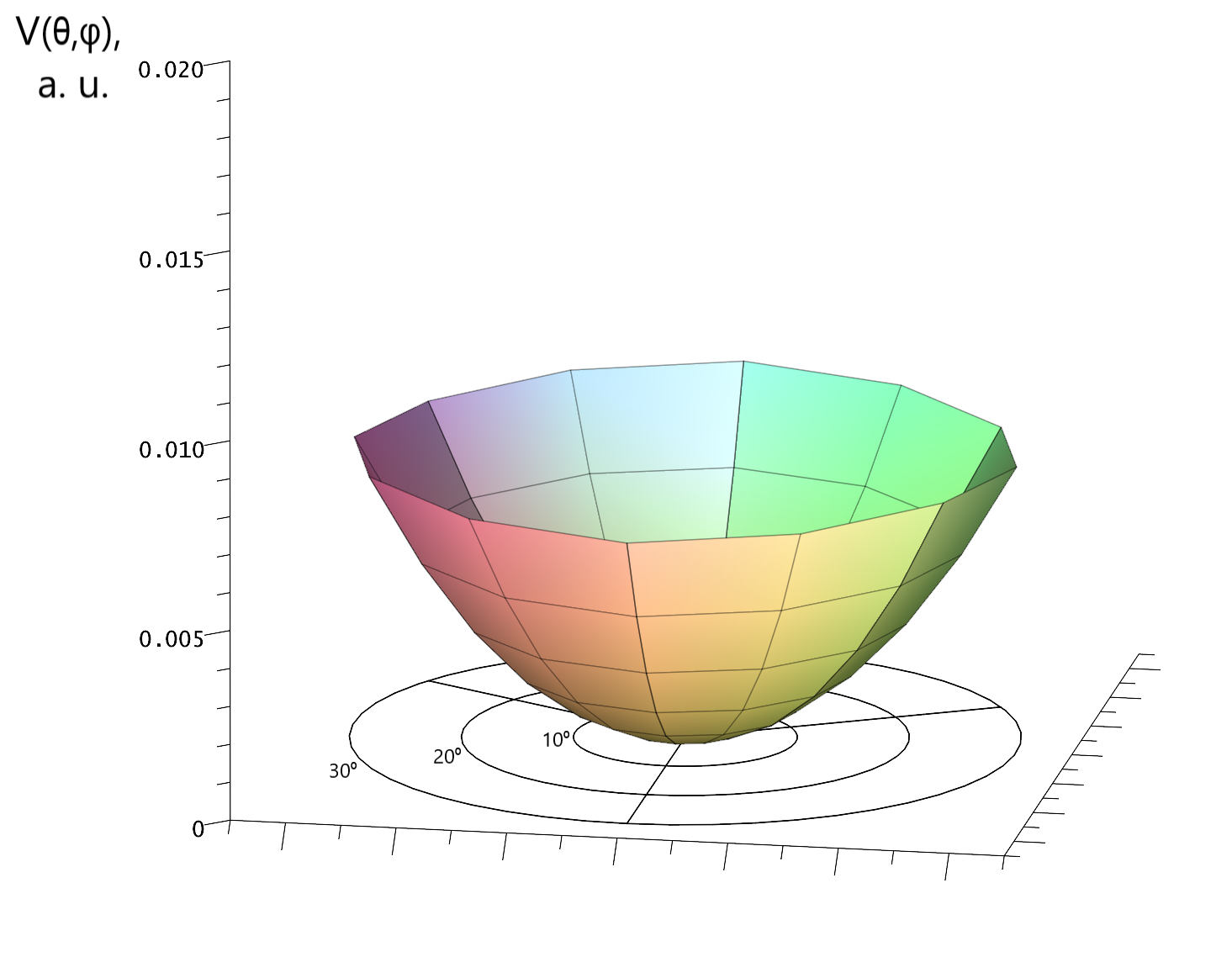}
  \caption{The angular dependence of the adiabatic potential at the equilibrium value $R=5.7\,\mathrm{a.u.}$ The azimuthal angle is $\varphi$, the radial coordinate is $\theta$. The sector-dividing lines correspond to the directions to the hydrogen atoms.}
  \label{fig:potential_azimuthal}
\end{figure}

The computed potential surface has a minimum near $R=5.7\, a.u.$ and $\theta=0$. The dependence on the angle $\varphi$ depicted on the Fig. \ref{fig:potential_azimuthal} becomes noticeable at large $\theta$. The difference between the energies for $\varphi=0^{\circ}$ and $\varphi=60^{\circ}$ for $\theta=30^\circ$ reaches $59.2 \,\mathrm{cm}^{-1}$ which constitutes $3.6\%$ of the absolute value of $V-V_{\rm eq}$. Not surprisingly it becomes stronger for smaller $R$ reaching $76.6\,\mathrm{cm}^{-1}$ ($11\%$ of the absolute value of $V-V_{\rm eq}$) for $R=5.5\, a.u.$. Nevertheless, the dependence on $R$ becomes significant only for $\theta\sim 30^\circ$, and our approximation for the $V(R,\theta,\varphi)-\bar{V}(R,\theta)$ not depending on $R$ is justified. The term $\delta V^{(2)}$ contributes at most $10^{-6}\mathrm{cm}^{-1}$ to the potential and can be neglected.

The harmonic approximation for the $\varphi$-averaged potential surface gives,
\begin{equation}
\omega_\parallel=345.17 \mathrm{cm}^{-1},\quad\omega_\perp=151.32 \mathrm{cm}^{-1}
\end{equation}
This may be compared with $\omega_{\text{Ra-O stretch}}=390.78\mathrm{cm}^{-1}$ and $\omega_{\text{Ra-O-C bend}}=164.96/168.68\mathrm{cm}^{-1}$ in \cite{yu2021probing} for $\,^{225}$RaOCH$_3^{+}$ ion.

\begin{figure}[h]
\centering
  \includegraphics[width=0.5\textwidth]{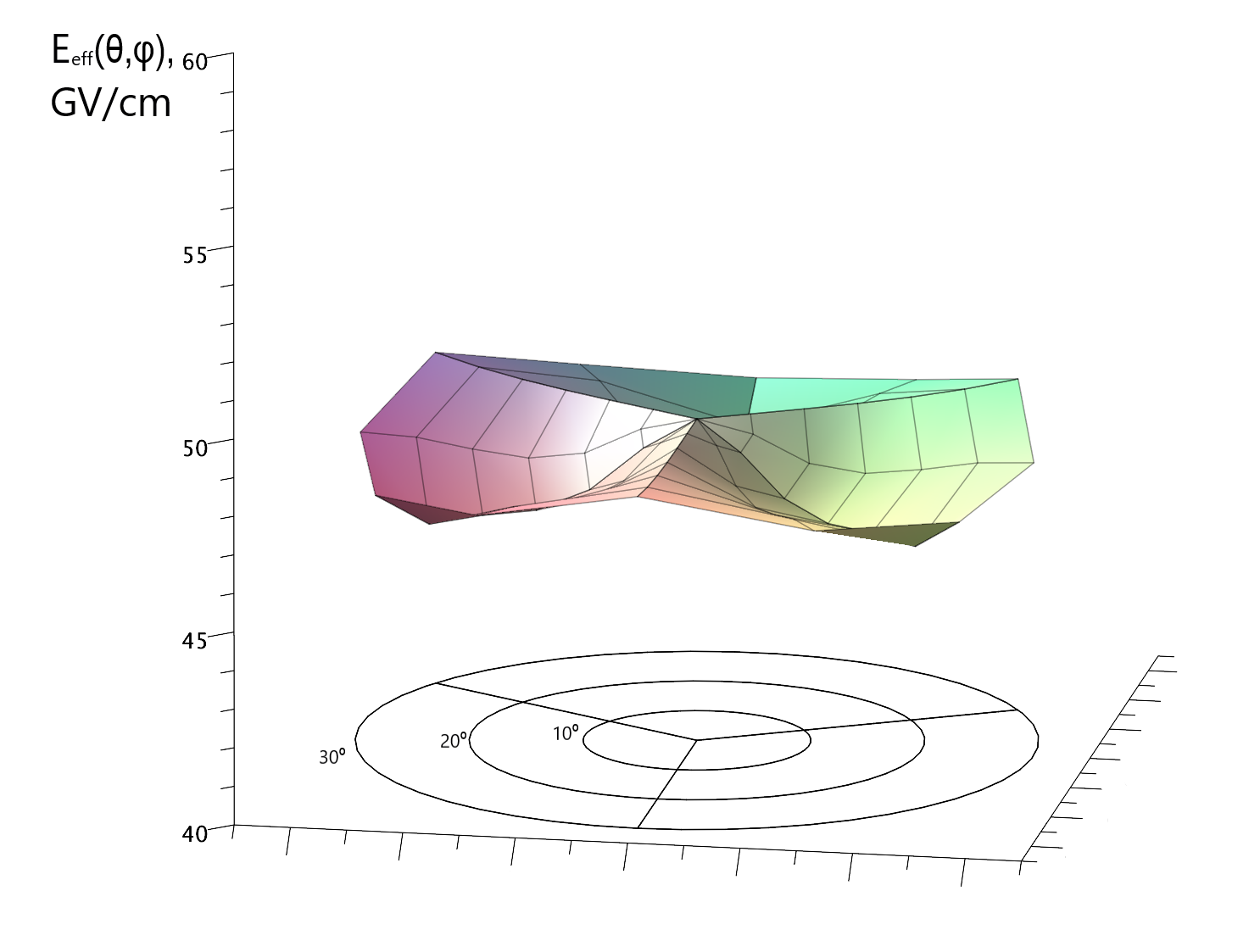}
  \caption{The angular dependence of the $E_{\rm eff}$ at $R=5.7\,\mathrm{a.u.}$}
  \label{fig:Eeff_azimuthal}
\end{figure}

\begin{figure}[h]
\centering
  \includegraphics[width=0.5\textwidth]{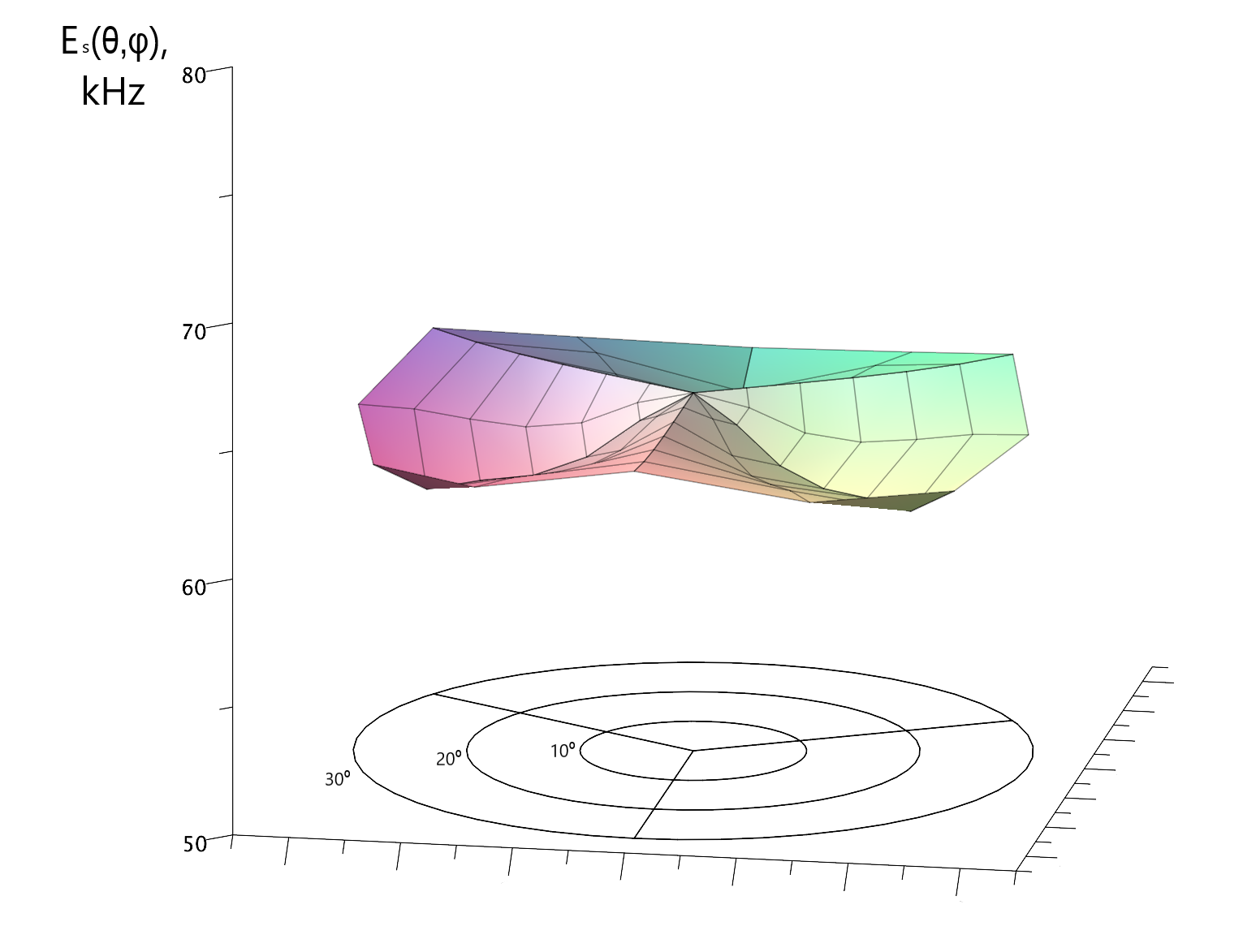}
  \caption{The angular dependence of the $E_{\rm s}$ at $R=5.7\,\mathrm{a.u.}$}
  \label{fig:Es_azimuthal}
\end{figure}

\begin{figure}[h]
\centering
  \includegraphics[width=0.5\textwidth]{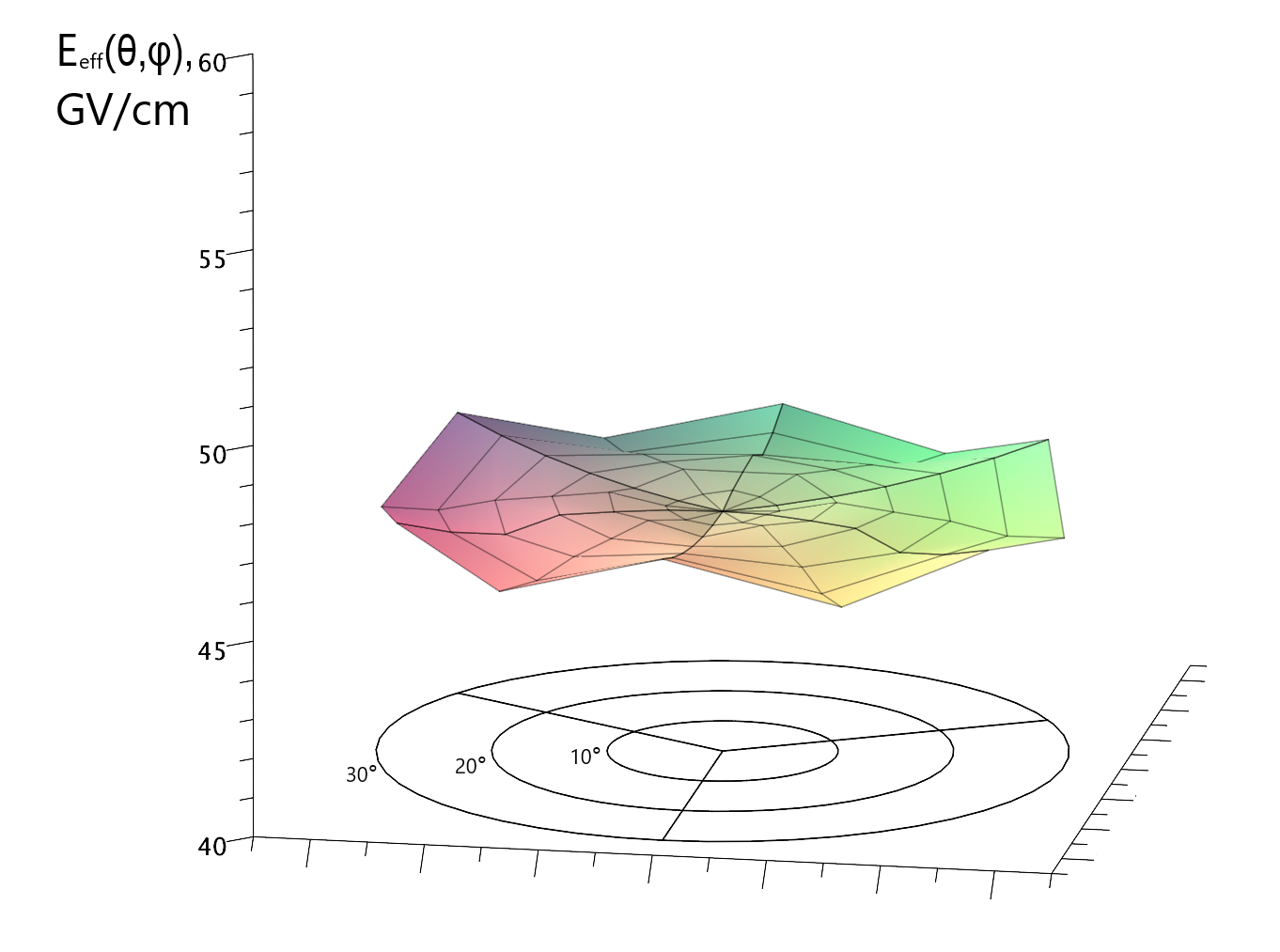}
  \caption{The angular dependence of the $E_{\rm eff}$ at $R=5.5\,\mathrm{a.u.}$}
  \label{fig:Eeff_azimuthal_55}
\end{figure}

\begin{figure}[h]
\centering
  \includegraphics[width=0.5\textwidth]{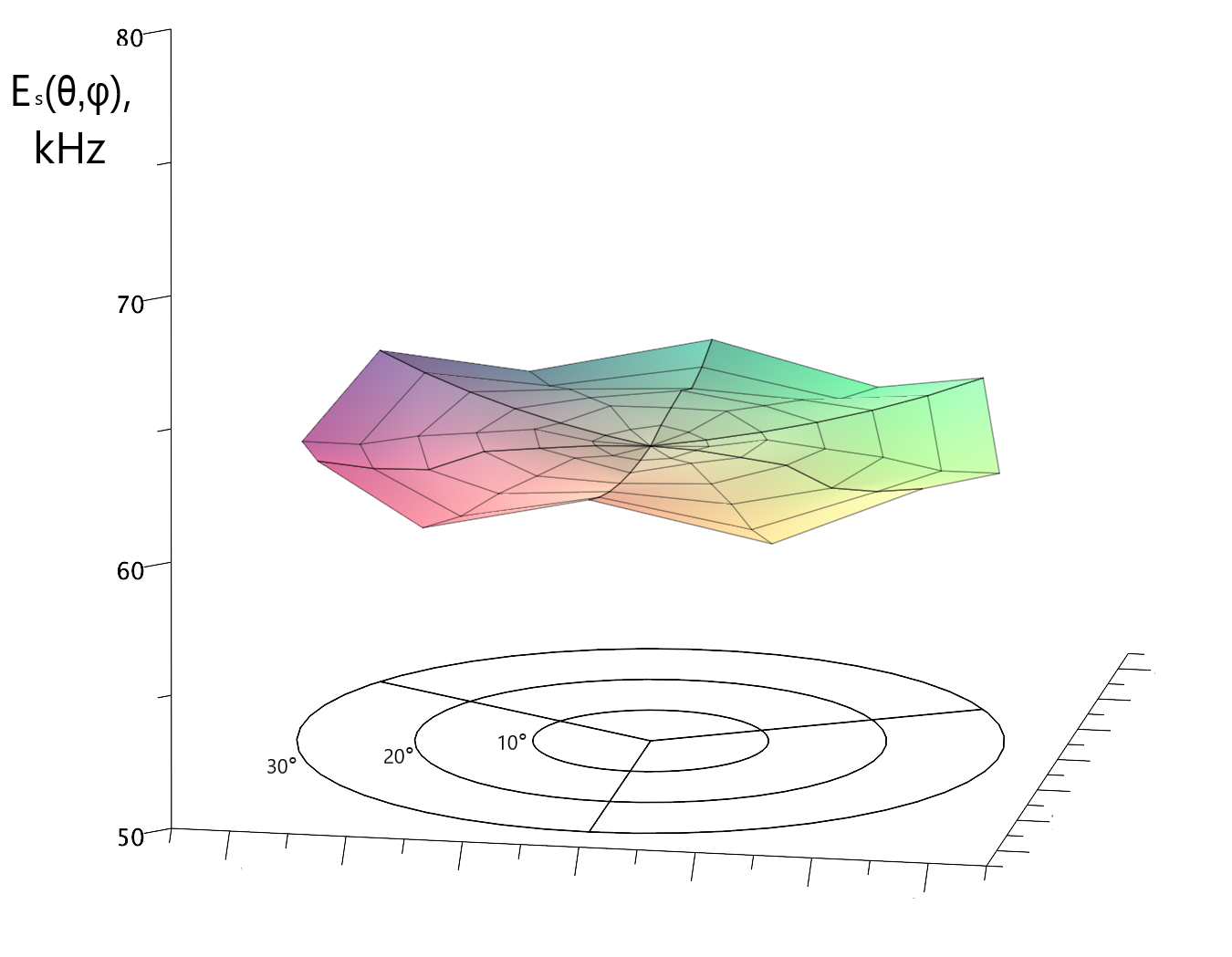}
  \caption{The angular dependence of the $E_{\rm s}$ at $R=5.5\,\mathrm{a.u.}$}
  \label{fig:Es_azimuthal_55}
\end{figure}

The dependence of the $\mathcal{P}$,$\mathcal{T}$-odd parameters on the angles $\theta$ and $\varphi$ is shown on the Fig. \ref{fig:Eeff_azimuthal} and Fig. \ref{fig:Es_azimuthal}. The dependence on $\varphi$ is somewhat smaller for $E_{\rm eff}$ in comparison with $E_{\rm s}$. At $R=5.7\,\mathrm{a.u.}$ and $\theta=30^\circ$ the difference between the values for $\varphi=0^\circ$ and $\varphi=60^\circ$ constitutes about $1\%$ for $E_{\rm eff}$ and $3\%$ for $E_{\rm s}$. For smaller $R$ the amplitude of the oscillations in $\varphi$ do not grow, instead they become more frequent as seen on the Fig. \ref{fig:Eeff_azimuthal_55} and on the Fig. \ref{fig:Es_azimuthal_55}.



We present the results for $\mathcal{P}$, $\mathcal{T}$-odd parameters both for the equilibrium configuration and for the rovibrational states in the Table \ref{tbl:tableResults}. For the lowest $K$-doublet with $v_\perp=0$ and $K=1$ the values are $E_{\rm eff}=47.647\,\mathrm{GV/cm}$ and $E_{\rm s}=62.109\,\mathrm{kHz}$. The results are close to the values obtained for other polar molecules with the Radium atom. This confirms the validity of our computational approach.

One can see that the relative difference between the equilibrium value and the value for the ground vibrational state of the RaOCH$_3$ molecule is larger than such difference for the excited vibrational states of the triatomic molecules RaOH and YbOH we studied earlier in \cite{ourRaOH,zakharova2021rovibrational,petrov2021sensitivity}. The primary role is played by the drop of sensitivity when the radium atom is bending in the direction between the H atoms that leads to the lowered value of the enhancement parameter averaged over $phi$ already for small $\theta$. Because $\theta$ may be considered the radial direction for the transverse vibrations $q_x$, $q_y$, already in the ground vibrational state the maximum contribution is given by $\theta\simeq 5^\circ$ and not by the equilibrium configuration. The fast drop with $\theta$ plays the main role in the difference between the equilibrium value and value for the $v=0$ state, with the $v=1$ state having almost the same enhancement parameter as the ground state. The effect is stronger for $E_{\rm s}$ parameter. The impact of the $\varphi$ dependence of the potential $V$ happen to be insignificant, amounting only to $10^{-4}\frac{\mathrm{GV}}{\mathrm{cm}}$ for $E_{\rm eff}$ and $10^4\,\mathrm{kHz}$ for $E_{\rm s}$. 

The centrifugal correction to the $\omega_\perp$ given by \eqref{omegaRot} only slightly changes the values of the $\mathcal{P}$,$\mathcal{T}$-odd parameters. As expected, it has a stronger influence on the $v_\perp=1$ state.

Because of the high computational costs we have not estimated the errors due to the basis selection and use of the effective potential. In this paper we also did not considered the anharmonicities of the potential that may impact the averaging over the longitudinal vibrations. The CCSD energy computations were done with convergence criterion $\Delta E\simeq 10^{-12}\,\mathrm{a.u.}$. The finite field computation error then may be estimated as $\Delta E_{\rm eff}\sim 10^{-5}\,\mathrm{GV/cm}$ and $\Delta E_{\rm s}\sim 10^{-5}\,\mathrm{kHz}$. The harmonic approximation error for the transverse vibrations is $\Delta\omega_\perp/\omega_\perp\simeq 0.7\%$ which is less than the centrifugal correction for $K>2$. This allows us to assume that our description of the dependence of the parameters on $v_\perp$ and $K$ is at least qualitatively right.

Our results stress the importance of the rovibrational effects for the computation of the symmetric top molecule sensitivity to the $\mathcal{P}$, $\mathcal{T}$-odd effects already within the harmonic approximation and for the ground vibrational state. We have taken into account the dependence of the potential on the bending direction $\varphi$. We have also considered the centrifugal and Coriolis effects associated with the rotation of the molecule around $Z$ axis. The impact of both effects happened to be very small. We will study the role of the anharmonicities and other couplings between the rotational and vibrational degrees of freedom in the future work.

\begin{table}[h]
\small
  \caption{The $\mathcal{P}$, $\mathcal{T}$-odd parameters for the equilibrium configuration and for the rovibrational states}
  \label{tbl:tableResults}
  \renewcommand{\arraystretch}{1.5}
  \begin{tabular*}{0.48\textwidth}{@{\extracolsep{\fill}}lllll}
    \hline\hline
& $v_\perp$ &  $K$ & $E_{\rm eff},\, \frac{\rm GV}{\rm cm}$ &$E_s,\, {\rm kHz}$\\
\hline
RaOCH$_3$ & \multicolumn{2}{l}{equilibrium}   &   48.346  & 64.015\\

RaOCH$_3$ & $v_\perp=0$ $l_v$ = 0 & $0$   &   47.930  & 63.436\\
& & $1$   &   47.929  & 63.435\\
& &  $2$   &   47.927  & 63.433\\
& & $3$   &   47.924  & 63.429\\
& & $4$   &   47.920  & 63.423\\

RaOCH$_3$  & $v_\perp=1$ $l_v$ = +1 & $0$   &   47.649  & 63.064\\
& & $1$   &   47.648  & 63.063\\
& & $2$   &   47.646  & 63.060\\
& & $3$   &   47.642  & 63.055\\
& & $4$   &   47.636  & 63.048\\
RaOCH$_3$  & $v_\perp=1$ $l_v$ = -1 & $0$   &   47.649  & 63.064\\
& & $1$   &   47.648  & 63.063\\
& & $2$   &   47.645  & 63.060\\
& & $3$   &   47.641  & 63.055\\
& & $4$   &   47.637  & 63.047\\

\hline
RaOCH$_3$\cite{zhang2021calculations} & \multicolumn{2}{l}{equilibrium}  & 54.2 & --\\
    \hline
RaOH\cite{ourRaOH} & \multicolumn{2}{l}{equilibrium}  & 48.866  & 64.788\\
& $v=1$ &      & 48.585 & 64.416\\
\hline
YbOH\cite{zakharova2021rovibrational} & \multicolumn{2}{l}{equilibrium} & 23.875 & 20.659 \\
& $v=1$ &
  & 23.576 & 20.548 \\
  \hline
 RaF \cite{kudashov2014ab} & & & 52.937 & 69.5 \\
 RaF \cite{gaul2019systematic} & & & 58.11 & 68.0\\
 YbF cGHF \cite{BergerYbF} & & & 24.0 & 20.6  \\
 YbF cGKS \cite{BergerYbF} & & & 19.6 & 16.9  \\
 ThO \cite{skripnikov2016combined} & & & 79.9 &  113.1\\
 HfF+ \cite{skripnikov2017communication} & & & 22.5 & 20.1 \\
 HfF+ \cite{fleig2017p} & & & 22.7 & 20.0 \\
 \hline\hline
  \end{tabular*}
\end{table}


\section{acknowledgement}

The work was supported by the Theoretical Physics and Mathematics Advancement Foundation “BASIS” (grant N\textsuperscript{\underline{o}} 21-1-5-72-1).
The author would like to thank A. Petrov, T. Isaev, and A. Titov for support.

\vfill

\bibliographystyle{apsrev}

\end{document}